\documentstyle[amssymb,epsf]{elsart}

\newtheorem{theorem}{Theorem}
\newtheorem{definition}{Definition}
\newcommand{\cs}[2]{\mbox{{\sf cshift-row}}_{#2}^{#1}}
\newcommand{\cc}[2]{\mbox{{\sf cshift-col}}_{#2}^{#1}}
\newcommand{\beq}{\begin{equation}}
\newcommand{\eeq}{\end{equation}}        
\newcommand{\bqa}{\begin{eqnarray}}        
\newcommand{\eqa}{\end{eqnarray}}        
\newcommand{\be}{\begin{enumerate}}
\newcommand{\ee}{\end{enumerate}}        
\newcommand{\bi}{\begin{itemize}}
\newcommand{\ei}{\end{itemize}}        
\newcommand{\fig}[1]{Fig.~\ref{#1}}
\renewcommand{\alg}[1]{Algorithm~\ref{#1}}

\renewcommand{\matrix}[1]{\mbox{\sf #1}}
\renewcommand{\vector}[1]{\mbox{\sf #1}}
\newcommand {\bmatrix}[1]{\mbox{\bf #1}}

\newcommand{\eg}{\mbox{\frenchspacing\em e.\hspace{0.4mm}g.{}}}
\newcommand{\ie}{\mbox{\frenchspacing\em i.\hspace{0.4mm}e.{}}}
\newcommand{\cf}{\mbox{\frenchspacing\em cf.{}}}
\newcommand{\vs}{\mbox{\frenchspacing\em vs.{}}}
\newcommand{\etc}{\mbox{\frenchspacing\em etc.{}}}
\newcommand{\mod}[1]{{\,\mbox{{\footnotesize mod}}_{#1}}}
\newcommand{\eq}[1]{{\frenchspacing Eq.~\ref{#1}}}

\newenvironment{algo}[2]{\vglue6pt\noindent
\begin{minipage}%
{\textwidth}
\begin{algorit}  #2 \rm #1\par \footnotesize%
\begin{center}%
\begin{minipage}{\textwidth}%
\begin{tabbing}}%
{\end{tabbing}%
\end{minipage}%
\end{center}%
\end{algorit}%
\end{minipage}}%
\newtheorem{algorit}{Algorithm}

\begin{document}

\begin{frontmatter}
  
  \title{\vspace*{-.5cm}
{\small\hfill HLRZ1998-59}\\
Hyper-Systolic Matrix Multiplication}
  
  \author[WUPW,HLRZ]{Th.~Lippert} \author[GRON]{N.~Petkov}
  \author[ENEA]{P.~Palazzari} \author[WUPW,HLRZ]{K.~Schilling}
  
  \address[WUPW]{Department of Physics, University of Wuppertal,
    D-42097 Wuppertal, Germany}

  \address[HLRZ]{HLRZ, c/o Research Center J\"{u}lich, D-52425
    J\"{u}lich, Germany}
  
  \address[GRON]{Institute of Mathematics and Computing Science,
    University of Groningen, PO Box 800, 9700 AV Groningen, The
    Netherlands}
  
  \address[ENEA]{ENEA, HPCN Project, C.\ R.\ Casaccia, Via
    Anguillarese, 301, S.P. 100, 00060 S.Maria di Galeria, Rome,
    Italy}
  
\begin{abstract}  
  A novel parallel algorithm for matrix multiplication is presented.
  The hyper-systolic algorithm makes use of a one-dimensional
  processor abstraction.  The procedure can be implemented on all
  types of parallel systems. It can handle matrix-vector
  multiplications as well as transposed matrix products.
\end{abstract}


\end{frontmatter}

\section{Introduction}
Matrix multiplication is a fundamental operation in most numerical
linear algebra applications.  Its efficient implementation on parallel
high performance computers, together with the implementation of other
basic linear algebra operations, is therefore an issue of prime
importance when providing these systems with scientific software
libraries \cite{DONGARRA}.  Consequently, considerable effort has been
devoted in the past to the development of efficient parallel matrix
multiplication algorithms, and this will remain a task in the future
as well.

A general rule, which applies not only to matrix multiplication is,
that the choice of a proper parallel algorithm strongly depends on the
architecture of the parallel computer on which the algorithm is to
run. System aspects, such as SIMD or MIMD mode of operation,
distributed or shared memory organization, cache or memory bank
structure, construction and latency of the communication network,
processor performance and size of local memory, etc., may render an
algorithm which is highly efficient for one system rather impractical
for another system. Even on a given system it may be necessary to
switch algorithms in different problem size domains.

As a consequence, one needs a diversity of algorithms for one and the
same operation, as well as systematic design approaches which allow to
construct new algorithms or to modify existing ones in such a way that
they suit both a given implementation system and problem size domain.
In the following, one such design approach is presented and applied to
matrix multiplication.  The procedure will lead us to a novel class of
parallel matrix multiplication algorithms which are applicable to
distributed memory computers whose interconnection pattern includes a
ring as a subset of the system connectivity. This novel scheme is
called the hyper-systolic matrix multiplication, as it is based on the
hyper-systolic parallel computing concept \cite{LIPPERT95}. The latter
is generalizable to any kind of commutative and associative operation
on abstract data types \cite{GALLI95}.  The communication complexity
of the hyper-systolic matrix multiplication is
$O(n^{2}\,p^{\frac{1}{2}})$, with $n$ being the matrix dimension and
$p$ the number of processors, and thus, it is comparable to best
parallel standard methods.

The work presented is part of a program to develop a {\em novel
  practicable} form of distributed BLAS-3 (PBLAS-3).  In a forthcoming
publication, we will give algorithms for transposed matrix products
and level-2 linear algebra computations.

The paper is organized as follows: in Section \ref{BACKGROUND}, we
shortly comment on systolic algorithms and the origin of
hyper-systolic algorithms and review in brief the problem of
distributed matrix multiplication. In Section \ref{SEC1}, we develop
the one-dimensional (1D) hyper-systolic matrix multiplication
algorithm in a systematic way, starting from Fox' algorithm on a
two-dimensional (2D) mesh \cite{FOX87}.  In Section \ref{SEC:HYPER},
the concept of the hyper-systolic algorithm involving two data arrays
is introduced. The algorithmic presentation of the hyper-systolic
matrix multiplication is given in Section \ref{MATPROD}, Section
\ref{BLOCKING} deals with the mapping of the problem onto parallel
systems and finally, Section \ref{TESTS} presents some results from
implementations on a SIMD computer.

\section{Background\label{BACKGROUND}}

\subsection{Systolic arrays} 

Systolic arrays are cellular automata models of parallel computing
structures in which data processing and transfer are pipelined and the
cells carry out functions of equal load between consecutive
communication events.  Systolic algorithms are parallel algorithms
which, as far as abstract automata models are concerned, make
efficient use of systolic arrays.  For more precise definitions of
systolic algorithms and arrays and for many examples, the reader is
referred to the monographs under references \cite{BOOK1} and
\cite{BOOK2} (for a number of systolic matrix multiplication
algorithms see Chapter 3 of \cite{BOOK2}).  The original motivation
behind the systolic array concept was its suitability for VLSI
implementation \cite{KUNG1,KUNG2}. Only a few systolic algorithms,
however, have been implemented in VLSI chips or hardware devices. With
the advent of commercially available parallel computers, systolic
algorithms have found an attractive implementation medium for they
match the local regular interconnection structure as present in or
easily implementable on many parallel architectures.

Systolic algorithms can efficiently be implemented in the SIMD model,
where---being synchronous and regular---they avoid time consuming
synchronization operations. Apart from these implementation issues, a
very attractive feature is the availability of methodologies for the
systematic design of systolic algorithms.  Projection of regular
dependence graphs has evolved as one such technique
\cite{BOOK1,BOOK2,CAPPELLO83,CAPPELLO87,QUINTON84,MOLDOVAN82,MOLDOVAN83,CLAUSS94,DARTE95,CLAUSS96}.

As shown elsewhere \cite{DONTJE91,PETKOV_FUZZY,PETKOV_FUZZY2}, such
structures can easily be transformed into data-parallel programs.  The
pattern of systolic arrays induces characteristic features into the
respective data-parallel programs. In particular, a data-parallel
program realizing a systolic algorithm consists of a sequence of
identical steps organized in a loop whose counter corresponds to the
clock of the underlying systolic array automaton model. Further, the
local regular interconnection pattern of a systolic array results in
the use of only local synchronized communication in the respective
data-parallel program as exemplified by the {\it shift}-type
operations (e.g. {\sf cshift} and {\sf eoshift}).

\subsection{Hyper-systolic algorithms}

Hyper-systolic algorithms have been introduced for various numerical
applications in order to further reduce the communication expense of
systolic algorithms.  The advantage of the hyper-systolic over the
systolic data flow first has been demonstrated for the case of
so-called $n^{2}$-problems, \ie, numerical problems that involve
$O(n^{2})$ computation events on pairs of elements in a system of $n$
elements \cite{LIPPERT95}.

The systolic computation of $n^{2}$-problems on a parallel computer
equipped with $p$ processors involves $O(np)$ communication events.
The hyper-systolic algorithm can reduce the communication complexity
to $O(np^{\frac{1}{2}})$, as has been shown for a prototype
$n^{2}$-problem, the computation of all $n^{2}$ two-body forces for a
system of $n$ gravitatively interacting bodies \cite{LIPPERT96}.  This
success makes us confident that hyper-systolic processing can be
applied to a variety of numerical problems which lead to $n^{2}$
computation events.  An important application is found in astrophysics
where the investigation of the dynamics and evolution of globular
clusters is of prime importance \cite{HEGGIE}.  Further examples of
applications are protein folding, polymer dynamics, polyelectrolytes,
global and local all-nearest-neighbours problems, genome analysis,
signal processing \etc\ \cite{LIPPERTPHD}.

Hyper-systolic algorithms are an extension of the systolic concept
\cite{LIPPERT95}.  Similar to systolic algorithms, data processing and
transfer are pipelined and the cells carry out functions of equal
load.  The three main differences are: {\em (i)} use of a changing
interconnection pattern throughout the execution of the algorithm,
{\em (ii)} use of multiple auxiliary data arrays for storage of
intermediate results, and {\em (iii)} the possible separation of
communication and computation.  The combination of these three
features leads to reduction of the communication overhead.

The use of a changing communication pattern is due to the
communication of data by different strides along a 1D ring in
different stages of a hyper-systolic algorithm. The regularity of the
communication pattern is retained, however.  As an example of a
regular but changing communication pattern one can think of an
algorithm in which each processor of a ring communicates data to its
first, second, fourth, \etc, neighbours in the first, second, \etc,
steps of the algorithm, respectively.

Auxiliary data arrays are needed for temporary storage of intermediate
results. In conventional systolic algorithms such results are
accumulated by shifting them from cell to cell. In the hyper-systolic
algorithm they are generated and accumulated in place for many cycles,
using multiple auxiliary data arrays, and are subsequently used to
compute the final results.

Use of a regular but changing communication pattern can be found in
some (conventional) systolic algorithms, such as the systolic
implementation \cite{BOOK2} of Eklund's matrix transposition algorithm
on a hyper-cube \cite{EKLUND}.  Use of multiple data arrays for
solving specific tasks, \eg\ problem partitioning, can also be found
in systolic algorithm literature (see Chapter 12 in \cite{EKLUND}.),
however, for purposes different of those aimed at in hyper-systolic
algorithms.  The unique combination of the two features mentioned
above in the hyper-systolic concept aims at achieving new quality:
substantial reduction of the communication overhead.

\subsection{Matrix multiplication on parallel systems}

In order to carry out matrix multiplications on distributed systems
one has to take care for communication efficiency, parallelism and
scalability of the implementation. For scalability of the
implementation it appears very important that the data layout chosen
is preserved in the course of the computation without need for
reordering of initial or result matrices.  Furthermore, the data flow
organization should not hinder the efficient usage of register, vector
or cache facilities that largely dominates the overall performance.  A
third requirement is the general applicability in linear algebra tasks
such as block factorization algorithms, where, additionally,
matrix-vector multiplication must be carried out
effectively\footnote{Unfortunately it often happens that the
  matrix-to-processor mapping, as chosen for optimal BLAS-3
  performance, fails in the case of BLAS-2 applications.}.  It is a
general observation that many parallel matrix multiplication
algorithms \cite{DEPREZ,PARCO95} meet the criteria of efficiency,
parallelism and scalability only partially.  For illustration, let us
for instance consider one of the favorite matrix multiplication
strategies, Cannon's algorithm \cite{CAN69}, applied to $n\times
n$-matrix multiplication, $\bmatrix{A}\times \bmatrix{B}$.

\begin{figure}[!htb]
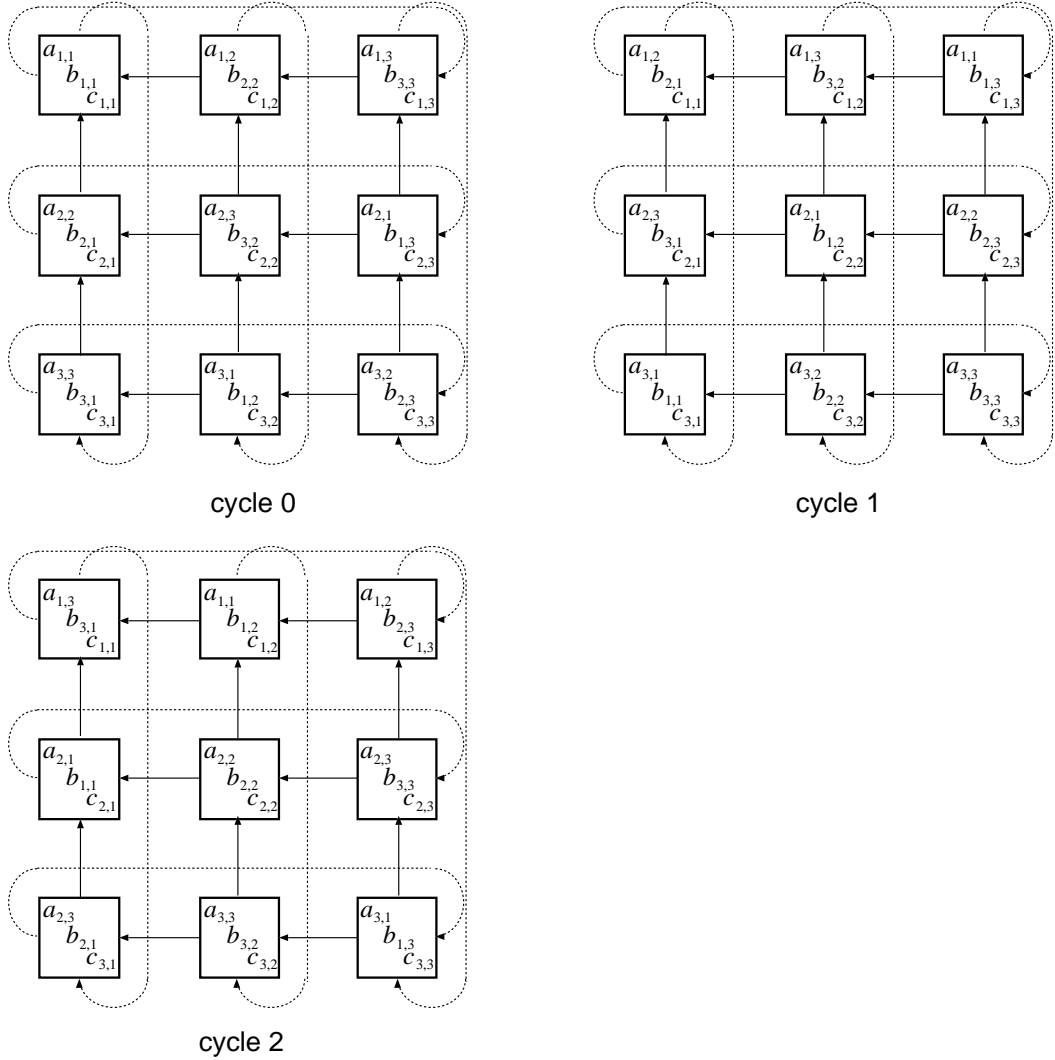

{\epsfxsize=\textwidth\epsfbox{cannon.eps}}
\vglue8pt
{\epsfxsize=.44\textwidth\epsfbox{cannon2.eps}}
\caption[Cannon's Algorithm]{
Systolic phase of Cannon's algorithm.
\label{CANNON2}}
\end{figure}

Elements ${a}_{i,k}$ and ${b}_{i,k}$ of the matrices $\matrix{A}$ and
$\matrix{B}$ are assigned to cells on a 2D grid, indexed by $i,j$.  In
a first step, Cannon's method involves a pre-skewing of $\matrix{A}$
and $\matrix{B}$ along the rows and columns, respectively.  Data
movement with different strides for each row/column is required, \cf\ 
\fig{CANNON1}. We note that this pre-skewing cannot be organized in
form of a regular communication pattern.
\begin{figure}[!htb]
\centerline{\epsfxsize=\textwidth\epsfbox{ini_cannon.eps}}
\caption[Preskewing of the Matrices]{
Preskewing of \matrix{A} and \matrix{B}.
\label{CANNON1}}
\end{figure}
The second systolic step of Cannon's method consists of circular
shifts of $\matrix{A}$ along the rows and of $\bmatrix{B}$ along the
columns, followed by the computation event, \cf\ \fig{CANNON2}. We
note that one can use blocks of elements as matrix entries.

The merit of Cannon's algorithm lies in its memory efficiency, as it
is possible to arrange the computation in such a way that no cell
holds more than one element (block) of each matrix. Disadvantages of
Cannon's algorithm are pre-skewing and the fact that the optimal data
layout for matrix multiplication requires one-to-all broadcast
operations applied to matrix-vector multiplication. Furthermore the
layout of the result vector differs from that of the input vector
\cite{KUMAR}.

\section{Design of a 1D Matrix Multiplication\label{SEC1}}

In this section, we develop a 1D hyper-systolic matrix multiplication
starting from a 2D algorithm that is related to the well known matrix
product scheme of Fox \cite{FOX87}.  In our systematic design approach
the 2D scheme will be transformed to a 1D systolic array
representation.

We have chosen a {\em skew ordering} as the fundamental representation
of the matrices $\matrix{A}$, $\matrix{B}$ and $\matrix{C}$.  As a
consequence, we will be able to carry out the computation fully in
parallel without even the requirement for indexed addressing
functionality of the target parallel computer.  Furthermore, no
reordering is required in the course of the computation because the
skew representation is preserved.

\subsection{Parallel Computational Problem} 
Given the $n\times m$ matrix $\bmatrix{A}$ and the $m\times n$ matrix
$\bmatrix{B}$, the matrix-matrix product reads \bqa c_{i,j} &=&
\sum_{k=1}^{m} a_{i,k}\, b_{k,j}, \quad i,j = 1, \dots,n.
\label{PRODUCT}
\eqa On a distributed memory machine, the elements of $\matrix{A}$ and
$\matrix{B}$ have to be spread appropriately across the different
processor nodes.

In our design approach, we will first show how to multiply $4\times 4$
matrices $\matrix{A}$ and $\matrix{B}$ on a 2D grid of nodes of size
$4\times 4$.  Then, the algorithm is mapped onto a 1D processing array
with $4$ nodes.

In Section \ref{MATPROD}, we present the hyper-systolic algorithm for
matrix multiplication of $p\times p$ matrices on an array of $p$
processors.

In Section \ref{BLOCKING}, we consider the general case \eq{PRODUCT}
of multiplying a $n\times m$ matrix $\bmatrix{A}$ and a $m\times n$
matrix $\bmatrix{B}$ on a 1D system of $p$ processors.  In order to
map the systolic model onto a parallel implementation machine we
choose hierarchy mapping with two different strategies, block and
cyclic assignment.

\subsection{2D Matrix Multiplication}
\subsubsection{Data alignment}
In the following we make use of the concept of abstract processor
arrays (APA) as defined in HPF \cite{HPF}.

\begin{figure}[!htb]
\centerline{\epsfxsize=.45\textwidth\epsfbox{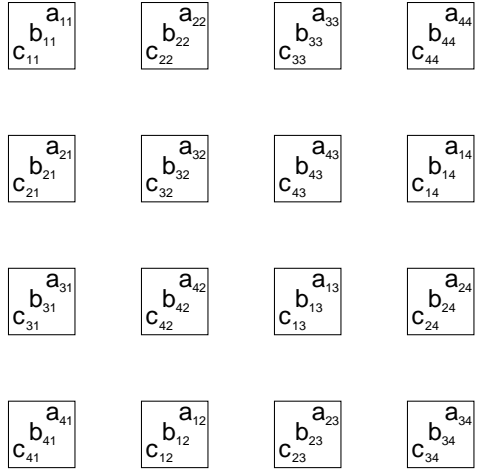}}
\caption[Template for a 2D Array Alignment]{
  Column-skewed distribution of the matrices $\matrix{A}$,
  $\matrix{B}$ and $\matrix{C}$ on a 2D APA.
\label{TEMPLATE}}
\end{figure}

The grid of boxes shown in \fig{TEMPLATE} is such a 2D APA on which
the matrices $\matrix{A}$, $\matrix{B}$ and $\matrix{C}$ involved in
the operation $\matrix{C}= \matrix{A} \matrix{B}$ are aligned in a
column-skewed fashion.

\subsubsection{Semi-systolic algorithm} 
Let the $p\times p$ matrices $\matrix{A}$, $\matrix{B}$ and
$\matrix{C}$ be distributed on a $p\times p$ processor array and let
for simplicity $p=4$.

\begin{figure}[!htb]
\centerline{\epsfxsize=\textwidth\epsfbox{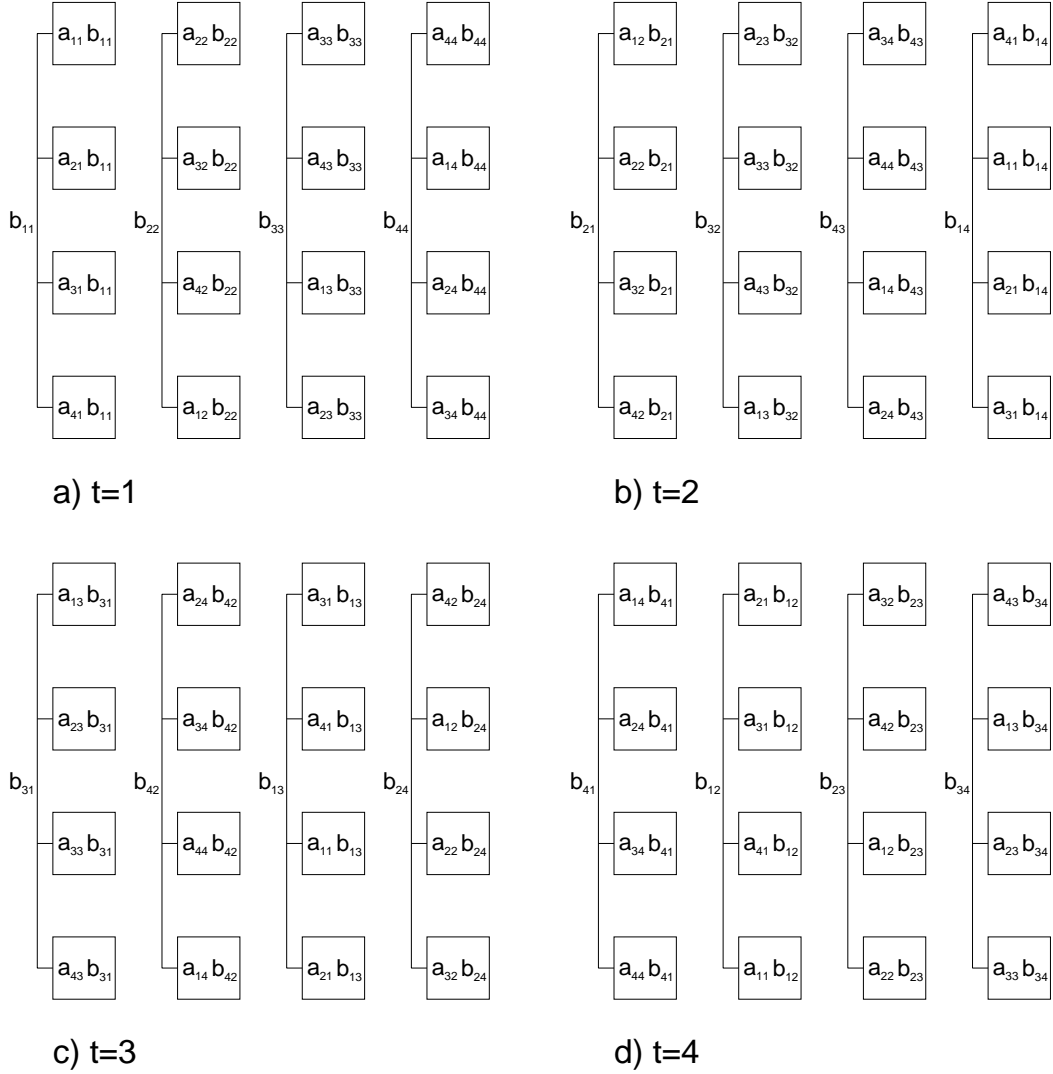}}
\caption{
Computation of partial products on a 2D APA.
\label{MODFOX}}
\end{figure}

The algorithm consists of $p$ (4) steps as follows: in the first step,
the matrix elements $b_{1,1}, b_{2,2},\dots,b_{p,p}$ of the matrix
$\matrix{B}$ in the first row of the APA are broadcast along the
corresponding columns and subsequently are multiplied with the
elements of the matrix $\matrix{A}$ in the respective columns of the
APA. The products (see \fig{MODFOX}a) are accumulated in the
corresponding elements of the matrix $\matrix{C}$, according to the
arrangement shown in \fig{TEMPLATE}.

At the end of the step, the elements of the matrix $\matrix{A}$ are
circularly shifted by one position to the left along the rows and by
one position downwards along the columns (compare the positions of the
elements of matrix $\matrix{A}$ in \fig{MODFOX}a and \fig{MODFOX}b).

In the second step, the processors of the second row of the APA
broadcast their corresponding elements of the matrix $\matrix{B}$ in
the respective columns of the APA, Fig.\ref{MODFOX}b. This operation
is followed by the same multiplication and accumulation operations and
circular shifting of the elements of the matrix $\matrix{A}$ as in the
first step.  

The algorithm proceeds with similar steps in which the processors of
the third, through $p$-th row of the array broadcast in turn their
elements of the matrix $\matrix{B}$, and the elements of matrix
$\matrix{A}$ are circularly shifted to the left and downwards by one
position, \cf\ Fig.\ref{MODFOX}c-d.  After a total number of $p$ such
steps all partial products which belong to the elements of the matrix
$\matrix{C}$ are accumulated in the corresponding processors.

The algorithm  is called semi-systolic as it involves a broadcast
operation of the row elements of $\matrix{B}$ in each step.  

\subsubsection{Semi-hyper-systolic algorithm}

\begin{figure}[!htb]
\centerline{\epsfxsize=.5\textwidth\epsfbox{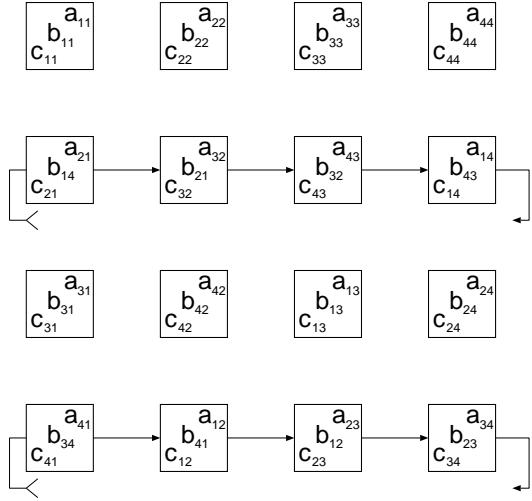}}
\caption[Initial Distribution for the Semi-Hyper-Systolic Matrix Product]{
  Initial distribution of data for the semi-hyper-systolic matrix
  product on a 2D APA.
\label{HYPERSTART}}
\end{figure}

We next consider the semi-hyper-systolic variant which corresponds to
the semi-systolic algorithm just described.  The initial distribution
of data is shown in \fig{HYPERSTART}. The distribution of the matrices
$\matrix{A}$ and $\matrix{C}$ is the same as the one shown in
\fig{TEMPLATE}.  The distribution of the matrix $\matrix{B}$ is
obtained from the distribution shown in \fig{TEMPLATE} by circular
shifting of the elements of $\matrix{B}$ in the $i$-th row of the
processor array by a stride of $(i-1)\mod{K}$, with $K=2$ for $p=4$.
In the particular example with $p=4$, the second and the fourth row of
$\matrix{B}$ are shifted by one position.

\begin{figure}[!htb]
  \centerline{\epsfxsize=\textwidth\epsfbox{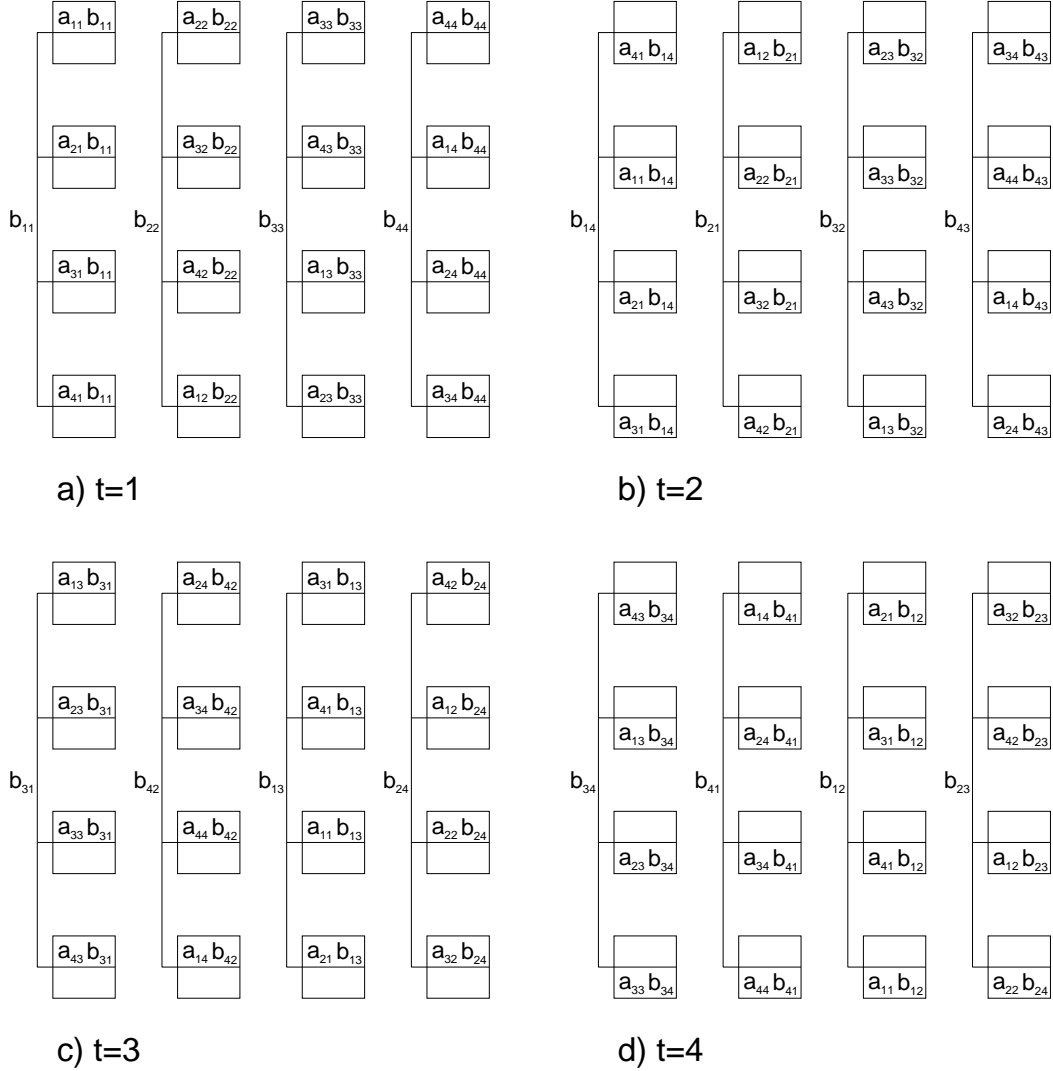}}
\caption[Semi-Hyper-Systolic Algorithm]{
  The products shown in the upper and the lower halves of the
  processors are accumulated in the auxiliary arrays
  $\matrix{C}^{(1)}$ and $\matrix{C}^{(2)}$.
\label{HYPER2D}}
\end{figure}

The algorithm consists of $p$ (4) steps as shown in \fig{HYPER2D}.  In
every step, each processor multiplies the elements of the matrix
$\matrix{A}$ with the elements of the matrix $\matrix{B}$ which it
receives via the associated broadcast line.  The partial product thus
computed is accumulated in one of two local variables.  The local
variables represent elements of two auxiliary arrays
$\matrix{C}^{(1)}$ and $\matrix{C}^{(2)}$ which distributed across the
processor array.  They are used to accumulate partial results for the
computations of the elements of the matrix $\matrix{C}$. Similar to
the original systolic algorithm the processors in the first, second,
\dots, $p$-th row broadcast the elements of the matrix $\matrix{B}$
they contain to all the processors of the corresponding columns in the
first, second, \dots, $p$-th step of the algorithm, respectively, see
\fig{HYPER2D}.

At the end of each step the matrix $\matrix{A}$ is cyclically shifted
downwards by one position.  Unlike the original algorithm the
horizontal shift of the matrix $\matrix{A}$ is performed {\em only
  every second} step by a stride of two elements.

The algorithm is completed by elemental addition of the auxiliary
arrays $\matrix{C}^{(1)}$ and $\matrix{C}^{(2)}$ which is preceded by
circular shifting of $\matrix{C}^{(2)}$ by one position to the left.

\subsubsection{Comparison to Cannon's algorithm} The 
semi-hyper-systolic algorithm defined above can be compared with
Cannon's algorithm.  Each of the two algorithms has merits and
shortcomings: Cannon's algorithm needs pre-skewing of the elements of
the matrix $\matrix{A}$ along the rows and the elements of matrix
$\matrix{B}$ along the columns of the 2D processor array which can be
considered as a disadvantage since it gives rise to additional
interprocessor communication that is not evenly distributed across
the processors.  On the other hand, the operations to be executed by
each processor in a given step of the algorithm are rather simple,
comprising one multiplication, one addition and two shift operations.
Also, Cannon's algorithm needs only local processor interconnections.

In contrast, the semi-hyper-systolic algorithm as illustrated in
\fig{HYPER2D} needs broadcasting lines and more complex control of the
operations which have to be carried out by the processors: in a given
step, each processor has to broadcast an element to all other
processors of the same column, and the matrix $\matrix{A}$ is
circularly shifted in two directions, where in horizontal direction a
stride of $2$ is required for every second step while in vertical
direction a stride of 1 is carried out in each step.  On the other
hand, this algorithm avoids pre-skewing of the matrices.

\subsection{Matrix Multiplication on a 1D Processor Array\label{DESIGN1D}}

Next we transform the semi-systolic algorithm discussed above into a
fully systolic one for a 1D processor array, in which a given
processor executes in sequence the tasks assigned so far to the
processors working on a given column of the 2D processor array.  This
procedure will eliminate the disadvantages of the semi-hyper-systolic
2D array algorithm since no broadcasting lines are needed, and
moreover, control structures become simpler.  Downward shifts of
$\matrix{A}$ merely amount to a re-assignment within the systolic cell
and do not involve actual communication between cells.  Note that the
application of such a transformation on the algorithm of Cannon will
not have the same effect as it would still be required to pre-skew one
of the matrices.

\subsubsection{Data alignment on a 1D processor array}

\begin{figure}[!htb]
\centerline{\epsfxsize=.45\textwidth\epsfbox{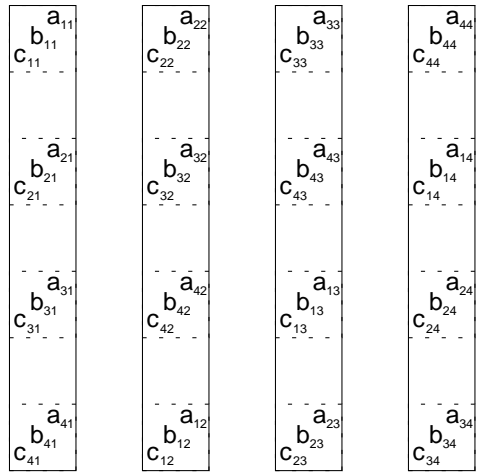}}
\caption[APA for a 1D systolic Array]{
  APA with 1D systolic array alignment.
\label{TEMPLATE2}}
\end{figure}

In the case of a 1D processor array, we assign each column of the 2D
array, \fig{TEMPLATE}, to one processor.  The resulting layout of the
matrices $\matrix{A}$, $\matrix{B}$ and $\matrix{C}$ is shown in
\fig{TEMPLATE2}.  The dotted lines indicate the location of the
elements in the cells of the previous 2D array.

\subsubsection{1D systolic algorithm}
Again we start with the systolic computation. The algorithm needs $p$
(4) steps. It does not require a broadcast of the matrix elements of
the matrix $\matrix{B}$.  In the first step, the matrix elements
$b_{1,1}, b_{2,2},\dots,b_{p,p}$ of the matrix $\matrix{B}$ which
reside in the first, second, \dots, $p$-th processor, respectively,
are multiplied with the elements of the respective columns of the
matrix $\matrix{A}$. The products (see \fig{MODFOX1D}a) are
accumulated in the corresponding elements of the matrix $\matrix{C}$,
according to the arrangement shown in \fig{TEMPLATE2}.  At the end of
the first step, the elements of the matrix $\matrix{A}$ are circularly
shifted by one position to the left along the rows.

\begin{figure}[!htb]
\centerline{\epsfxsize=\textwidth\epsfbox{modfox1d.eps}}
\caption[APA for Systolic Matrix Product]{
  APA for the systolic computation of partial products on a
  !D array.
\label{MODFOX1D}}
\end{figure}

In the second step, the second row of the matrix $\matrix{B}$ is
involved, see Fig.~\ref{MODFOX1D}b. Its elements are multiplied by the
elements of \matrix{A} and the partial products are accumulated in
their proper locations, \ie, in the second step the products are
copied to the elements of $\matrix{C}$ which are circularly assigned
one position downwards.  Subsequently, $\matrix{A}$ is circularly
shifted to the left by one position.

The algorithm proceeds with similar steps in which the elements of the
third, through $p$-th row the matrix $\matrix{B}$ are multiplied with
the elements of $\matrix{A}$, the products being assigned to a row of
$\matrix{C}$ at locations $i-1$ positions downwards the row in the
$i$-th step, with the elements of matrix $\matrix{A}$ being circularly
shifted to the left, \cf\ Fig.~\ref{MODFOX1D}c-d.  After a total
number of $p$ (4) steps all partial products which belong to the elements
of the matrix $\matrix{C}$ are accumulated in the corresponding
processors.

\subsubsection{1D hyper-systolic algorithm}

\begin{figure}[!htb]
\centerline{\epsfxsize=.5\textwidth\epsfbox{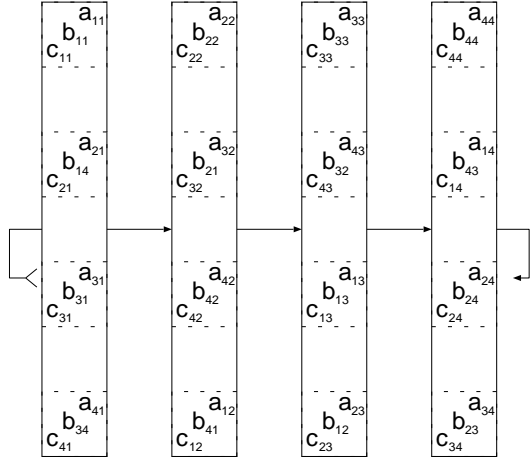}}
\caption[Initial Distribution of the Matrix Product]{
  Initial distribution of data for the hyper-systolic matrix product
  on a 1D APA.
\label{HYPERSTART1D}}
\end{figure}

Let us turn now to the 1D realization of the hyper-systolic algorithm.
The initial distribution---as in the 2D case---requires a partial
skewing of $\matrix{B}$ along the rows, see \fig{HYPERSTART1D}.  The
distribution of the matrix $\matrix{B}$ is obtained from the
distribution shown in \fig{TEMPLATE2} by circular shifting of the
elements of $\matrix{B}$ in the $i$-th row of the processor array by a
stride of $(i-1)\mod{K}$, with $K=2$ for $p=4$.

\begin{figure}[!htb]
  \centerline{\epsfxsize=\textwidth\epsfbox{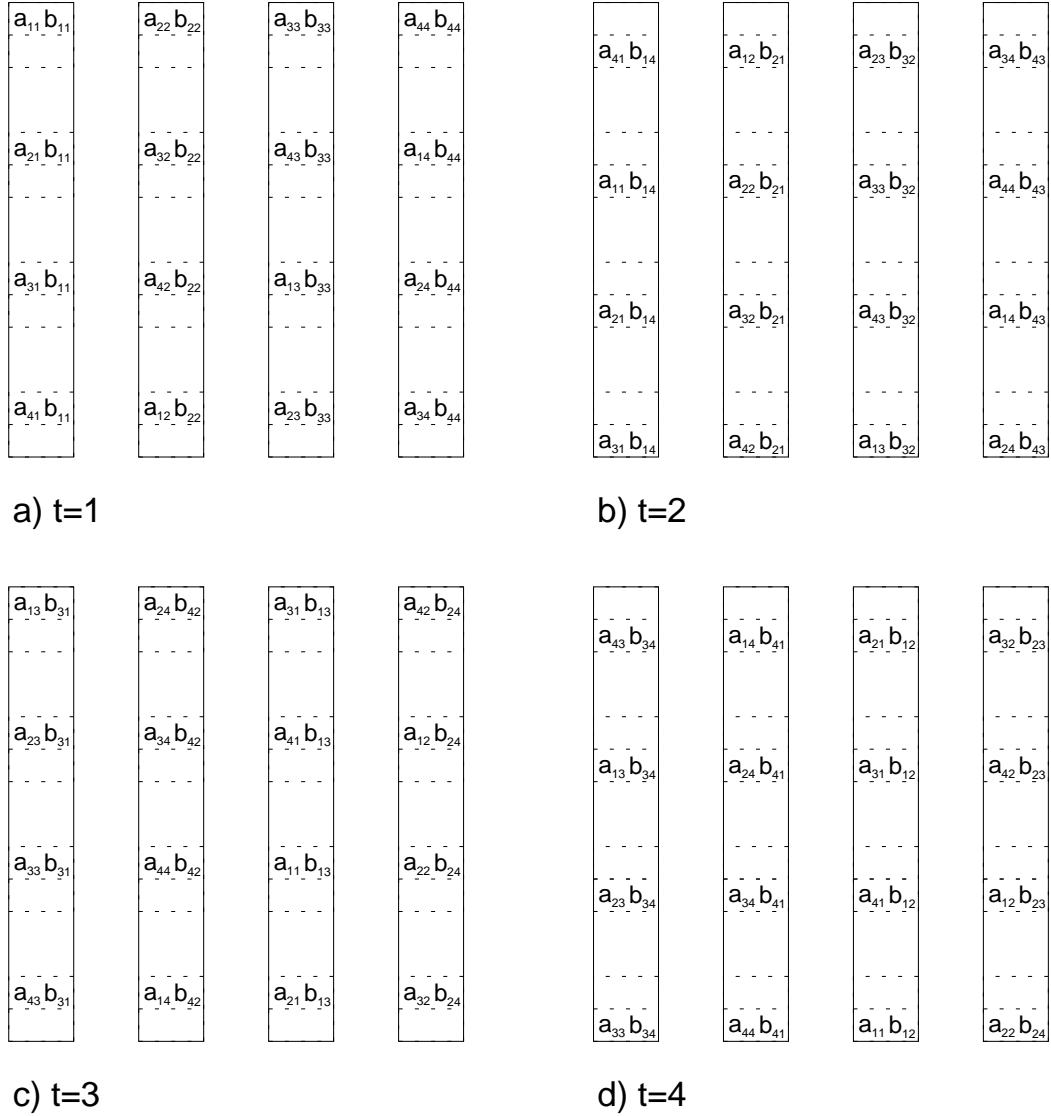}}
\caption[Hyper-Systolic Algorithm for the Matrix Product]{
Hyper-systolic algorithm on a 1D array.
\label{HYPER}}
\end{figure}

Step by step each processor multiplies the element of the matrix
$\matrix{A}$ it contains with the corresponding element of the matrix
$\matrix{B}$. The partial product thus computed is accumulated in one
of two local variables ($K=2$).  The products computed in alternate
steps are again stored alternately in one of the two local variables,
in an analogous fashion as for the systolic algorithm, \ie, in the
$i$-th step, the product is assigned to a row located $i-1$ elements
downwards, \cf\ \fig{HYPER}.
We emphasize again that $\matrix{A}$ is shifted only every second step
in horizontal direction to the left by two elements.  For the general
case with $p$ processors, we have to carry a shift by $\tilde K$
elements in every $K$-th step.

The algorithm is completed by elemental addition of the auxiliary
arrays $\matrix{C}^{(1)}$ and $\matrix{C}^{(2)}$ which is preceded by
circular shifting of $\matrix{C}^{(2)}$ by one position to the left.

\subsubsection{Summary}

In a systematic design approach, we developed a matrix multiplication
algorithm on a 1D abstract processor array, starting from an algorithm
on a 2D APA.  The algorithmic transformations led to a 1D
hyper-systolic scheme
\begin{itemize}
\item that avoids broadcast lines required in the 2D case,
\item that, given $p$ processors, shows a complexity of interprocessor
  communication on the 1D APA which is equal to that of Cannon's 2D
  algorithm,
\item that avoids skewing operations and reordering.
\end{itemize}
So far, we illustrated the scheme using $4\times 4$ matrices
distributed column-wise over 4 systolic cells.  In section
\ref{MATPROD}, we generalize the $4\times 4$ problem to a $p\times p$
system, distributed on a $p$ processor array. In the case of the
$4\times 4$ matrices, the shift constant was 2. For the general case,
we shall introduce a stride $K$ which is carried out $\tilde K$ times,
where the integer constants $K$ and $\tilde K$ fulfil $K\tilde K=p$.

\section{Hyper-Systolic Bases\label{SEC:HYPER}}

Before we present the general hyper-systolic matrix product, we define
the hyper-systolic algorithm for two data streams. Such a situation is
given in the computation of convolution and correlation problems.  The
scheme will later be adapted to the matrix multiplication
\cite{LIPPERTPHD}.

\subsection{Hyper-Systolic Recipe}
Let $\vector{x}$ and $\vector{z}$ be two 1D arrays both of length
$n$.  Assume that functions
\beq%
F_{i}=\bigoplus_{j=1}^{n}
f(x_i,z_j) 
\eeq%
for each $i$ have to be computed, with $\oplus$ being an associative
and commutative operator.  The computation can readily be carried out
by usage of systolic algorithms on a ring of systolic cells
\cite{LIPPERT96}.  However, one can observe redundant interprocessor
communication in this process that can be removed in hyper-systolic
processing.  The general recipe to find optimal hyper-systolic bases
reads:

Let the numerical problem be computable on two 1D systolic processor
arrays by use of a 1D systolic algorithm.  Let the two 1D data streams
$\vector{x}$ and $\vector{z}$ both of length $n$ be mapped onto themselves by
two sequences of (circular) shifts on the systolic array.  The
hyper-systolic scheme to compute the problem is constructed as
follows:
\begin{enumerate}
\item For the systolic array $\vector{x}$ of length $n$, $k$ replicas are
  generated by shifting the array $\vector{x}$ $k$ times by strides
  $a_{t}$, $1\le t\le k$, where all shifted arrays are stored as
  intermediate elements on the cells as arrays $\hat{\vector{x}_{t}}$,
  $1\le t\le k$.
    
  For the systolic array $\vector{z}$ of length $n$, $k'$ replicas are
  generated by shifting the array $\vector{x}$ $k'$ times by strides
  $b_{t'}$, $1\le t'\le k'$, where all shifted arrays are stored as
  intermediate elements on the cells as arrays $\hat{\vector{z}_{t'}}$,
  $1\le t'\le k'$.
  \item The sequences of strides $\{ a_{t}\} $, $1\le t\le k$ and $\{
    b_{t'}\} $, $1\le t'\le k'$ are determined such that
    \begin{enumerate}
    \item all pairings of data elements are present at least once,
    \item the total communication cost is minimized.
    \end{enumerate}
  \item After each communication event the computations can be carried
    out and the results are assigned to the corresponding intermediate
    result arrays $\hat{\vector{y}}_{t}$, $1\le t\le k+1$.  If elements
    occur more than once they are accounted for by a multiplicity
    table in order to avoid multiple counting.
  \item One collector array $\vector{y}$ is moved by strides that follow
    the inverse of the sequence $A_k$ of strides of the initial phase.
    In each step of the back-shift phase the required intermediate
    result arrays $\hat{\vector{y}}_{t}$ are added to $\vector{y}$.
  \end{enumerate}

\subsection{The Hyper-Systolic Optimization Problem}
Parallel machines support logical 1D chains of processors in form of
linear arrays or rings. However, circular shifts along the 1D ring in
general lead to different hardware communication expenses for
different strides.

The optimal sequence of strides for minimal interprocessor
communication will depend on the interprocessor communication cost for
a given stride.  In order to minimize the communication cost effect on
a given machine, we introduce a cost function ${ C}(a_{i})$, as a
function of the stride $a_{i}$.

For the sake of argumentation, let us first assume the costs of
communication for each array $\vector{x}$ and $\vector{z}$ on the systolic
ring to be constant for any stride $a_{i}$, $b_{i}$.
$C(a_i)=C(b_{\hat i})=\mbox{const}$.
\begin{definition} {\rm\bf - Optimization Problem for $C(a_i)=C(b_{\hat i})=\mbox{const.}$ }\\
\label{DEF:OPTI2}
Let $I$ be the set of integers $m=\{ 0,1,2\dots,n-1\}\in
{\Bbb{N}}^n_0, n\in {\Bbb{N}}$. Find the two ordered multi-sets
$A_k=(a_0=0,a_1,a_2,a_3,\dots,a_k)\in {\Bbb{N}}^{k+1}_0$ of $k+1$
integers and $B_{k'}=(b_0=0,b_1,b_2,b_3,\dots,b_{k'})\in
{\Bbb{N}}^{K+1}_0$ of $k'+1$ integers, with $k+k'$ being a minimum,
where each $m\in I$, ($0\le m\le n-1$), can be represented at least
once as the sum of two ordered partial sums
  \beq%
  m=(a_i+a_{i+1}+\dots +a_{i+j})+ (b_{\hat i}+b_{{\hat i}+1}+\dots
  +b_{{\hat i}+\hat j}),
  \label{EQ:ADD2}
  \eeq%
  with \beq 0\le i+j\le k,\qquad i,j \in {\Bbb{N}}_0,\qquad 0\le \hat
  i+\hat j\le k',\qquad \hat i,\hat j \in {\Bbb{N}}_0.
  \label{EQ:ADDII2}
  \eeq%
\end{definition}

\paragraph{Lower bound on $k+k'$}
A lower bound for the minimal number of non-zero elements of $A_{k}$
can be derived that will deliver optimal complexity.
\begin{theorem}\label{THE:MINIMUM2}
  Let $A_k$ and $B_{k'}$ be two bases solving the optimization problem for the
  hyper-systolic algorithm with 2 arrays.  Then the minimal length $k+k'$
  is given by
\beq%
k=k'=\sqrt{n}-1
  \label{EQ:MINIMAL2}.
\eeq%
\end{theorem}
\begin{pf}
  The total number of combinations required is $n^2$ as each element of the
  first array must come into contact with the $n$ elements of the second
  array.  Let the matrices ${\cal H}_1$ and ${\cal H}_2$ be realized by
  $k-1$ and $k'-1$ shifts respectively. In that case each element of the
  first matrix can be combined with $k'$ elements of the second matrix,
  therefore the possible number of combinations will be $nkk'$. Given
  $n=kk'$, the minimum number of circular shifts $k+k'$ is attained for
  $k=k'=\sqrt{n}-1$.\hfill\qed
\end{pf}

Therefore, the complexity for the interprocessor communication of a
hyper-systo\-lic algorithm for $C(a_{i})=C(b_i)=\mbox{const.}$ is
bounded from below by $3(\sqrt{n}-1)$ shifts, where we have already
included the costs for the back-shifts.

\subsection{${C}(a_{i})$ and  ${C}(b_{\hat i})\ne \mbox{const}$}
We now assume that the cost for a circular shift is a function of the
strides $a_{i}$.  The optimization problem of definition
\ref{DEF:OPTI2} is modified only slightly, however, the construction
of an optimal base can be quite complicated.
\begin{definition} {\rm\bf - Optimization Problem for $C(a_i)=C(b_{\hat
      i})\ne\mbox{const.}$ }\\
\label{DEF:OPTII2}
  Let $I$ be the set of integers $m=\{ 0,1,2\dots,n-1\}\in
  {\Bbb{N}}^n_0, n\in {\Bbb{N}}$. Find the two ordered multi-sets
  $A_k=(a_0=0,a_1,a_2,a_3,\dots,a_k)\in {\Bbb{N}}^{k+1}_0$ of $k+1$
  integers and $B_{k'}=(b_0=0,b_1,b_2,b_3,\dots,b_{k'})\in
  {\Bbb{N}}^{K+1}_0$ of $k'+1$ integers, with the total cost
\beq%
C_{\mbox{\footnotesize total}}=\sum_{i=1}^k C(a_i)+\sum_{\hat i=1}^{k'}
C(b_{\hat i})
\eeq%
being a minimum, where each $m\in I$, ($0\le m\le n-1$), can be
represented at least once as the sum of two ordered partial sums
  \beq%
  m=(a_i+a_{i+1}+\dots +a_{i+j})+ (b_{\hat i}+b_{{\hat i}+1}+\dots
  +b_{{\hat i}+\hat j}),
  \label{EQ:ADDA2}
  \eeq%
  with \beq 0\le i+j\le k,\qquad i,j \in {\Bbb{N}}_0,\qquad \hat
  i+\hat j\le k',\qquad \hat i,\hat j \in {\Bbb{N}}_0.
  \label{EQ:ADDAII2}
  \eeq%
\end{definition}

\subsection{Regular Bases}

The $4\times 4$ problem presented in Section \ref{SEC1} uses so-called
regular bases.  This prescription turns out to be optimal for equal
cost of any stride executed in circular shift operations on the ring.
Regular hyper-systolic bases are advantageous as they require only two
distinct strides.

\begin{definition} {\rm\bf - Regular Bases}
\label{DEF:REGULARBASE2}
  The regular bases are given by
  \beq%
  \renewcommand{\arraystretch}{.05}
  \arraycolsep=0pt
  \begin{array}{ccccccccc}
    A_{k=K-1} &:=& \Big(0, & 1,1,\dots,1\Big) \\
    & &       & \underbrace{\makebox[1.2cm]{}}& \\
    & &       & K - 1\\
  \end{array},
\label{EQ:HSBASEA}\quad
  \renewcommand{\arraystretch}{.05}
  \arraycolsep=0pt
  \begin{array}{ccccccccc}
    B_{k'=\tilde K-1} &:=& \Big(0, & K,K,\dots,K\Big) \\
    & &       & \underbrace{\makebox[1.6cm]{}}& \\
    & &       & \tilde K - 1\\
  \end{array}, \quad  K\times\tilde K =n.
  \eeq%
\end{definition}

The completeness of a base pair is defined in terms of the $h$-range
of the base, a notion borrowed from additive number theory
\cite{HOFMEISTER,UNSOLVED}:
\begin{theorem}
  The $h$-range of a regular base is $n$.
\end{theorem}
\begin{pf} Let
  \begin{equation}
    r:=m\mod{\tilde K}\rightarrow r< K,
  \end{equation}
  as $K\tilde K=n$.  There are $K-1$ elements $a_{i}=1\in A_{k}$.
  Thus any $r$ with $0\le r\le K-1$ $r\in{\Bbb N}_{0}$ can be
  represented as partial sum by the elements $a_{i}=1$, $a_{i}\in
  A_{k}$.  The partial sums of $B_{k'}$,
  \begin{equation}
   \sum_{l=i}^j b_l=
   K\sum_{l=i}^j 1 = (j-i+1) K<n, 
  \end{equation}
  are integer multiples of K.  Adding the partial sums to $r$ we can
  therefore represent any element $m\in I$. Thus, the $h$-range of the
  base pair $A_{k=K-1}$ and $B_{k'=\tilde K-1}$ is $n$, \ie, the base
  pair is complete.\hfill\qed
\end{pf} 
\begin{theorem}
  The lower bound to the minimal length of the regular bases for a
  given $h$-range $n$ is $K=\tilde K=\sqrt{n}$.
\end{theorem}
\begin{pf}
  The regular base $A_{k}$ is complete. 
  \begin{equation}
    k=K+\tilde K-1 \rightarrow
    k=K+\frac{n}{K}-1.
  \end{equation}
  Differentiation gives $K=\sqrt{n}$.\hfill\qed
\end{pf}

\begin{theorem}\label{THE:GAIN2}
  The communication gain factor $R$ that compares the regular
  hyper-systolic to the systolic algorithm is:
\beq%
R=\frac{n-1}{2K+\tilde K-3}
\approx \frac{\sqrt{n}}{3}.
\label{EQ:RATIO2ARRAY}
\eeq%
\end{theorem}
\begin{pf}
  Let $K=\tilde K=\sqrt{n}$.  One needs $K-1$ shifts by $1$ and
  $\tilde K-1$ shifts by $K$ in forward direction and again $K-1$
  shifts by $1$ in backward direction, respectively; therefore, the
  total number of shifts required turns out to be
\beq%
T=(2K+\tilde K-3).
\label{EQ:COMPLEXR}
\eeq%
The standard systolic computation requires $n-1$ shifts altogether.\hfill\qed
\end{pf}

\section{Hyper-Systolic Matrix Product\label{MATPROD}}
Next we present the general formulation of the systolic and
hyper-systolic matrix product in terms of a pseudo-code formulation.
The size of the matrices is $p\times p$ and the 1D processor array
consists of $p$ nodes.

\subsection{Systolic Algorithm}

The systolic version of the matrix product of two matrices $\matrix{A}$ and
$\matrix{B}$ is given in \alg{SYSTOLIC}. The matrices are represented
in skew order.

\begin{algo}{Systolic matrix-matrix multiplication.}%
            {\label{SYSTOLIC}}
for\= for\= for\= for\= for\= for\= \kill  
foreach processor $i=1:p$ $\in$ systolic array                  \\
\> for $j=1:p$                    \\
\> \> for $l=1:p$               \\ 
\> \> \>  $c_{l,i}=c_{l,i}+\cc{1-j}{p}(a_{l,i})\, b_{j,i}$  \\
\> \> end for                     \\ 
\> \> for $k=1:p$                 \\
\> \> \> $a_{k,i}=\cs{1}{p}(a_{k,i})$\\
\> \> end for                     \\
\> end for                        \\ 
end foreach                       \\ 
\end{algo}

We have simplified the representation of data movements and
assignments by introduction of two functions:

{\sf\bf cshift-row:} horizontal circular shift of data by a stride of
$k$ on a ring of cells numbered from $1$ to $n$, {\sf cshift-row}
involves interprocessor communication.
\begin{equation}
  \cs{k}{n}(a_{j,i}):=a_{j,(i+k-1+n)\mod{n}+1}.
\label{EQ:CSHIFT1}
\end{equation}

{\sf\bf cshift-col:} vertical circular shift by a stride of $k$ for
the vector of $n$ elements within the systolic cells. {\sf cshift-col} only amounts to memory assignments.
\begin{equation}
\cc{k}{n}(a_{j,i}):=a_{(j+k-1+n)\mod{n}+1,i}.
\end{equation}

The algorithm is completely regular. Each cell executes one compute
operation together with an assignment followed by a circular shift of
the matrix $\matrix{A}$ in each systolic cycle.  The skew order is not
destroyed during execution of the algorithm.  Note that for each
processor inner cell assignment operations ({\sf col-cshift}) are
executed using equal strides in a given step of the parallel
algorithm. Hence, one global address suffices and further address
computations are not required!

\subsection{Hyper-systolic Algorithm}
It has been already noticed in section \ref{DESIGN1D}, that the
complexity of the systolic computation is not competitive with
Cannon's algorithm. However, the hyper-systolic computation belongs to the same
complexity class as Cannon's algorithm. It is fully pipelined and
parallel, and does not require any skewing steps to align or re-align
matrix elements.

\subsubsection{Regular Bases}

We employ the regular bases constructed for the hyper-systolic system.
We add a third base $C$ to account for the back shifts:
\beq%
  \begin{array}{rrrrrrrrrrrrrrrrrrrr}
    A_{k=\tilde K-1} &= \Big(& 0,&  K,& K,&\dots, K&\Big) \\
    B_{k'= K-1}      &= \Big(& 0,& -1,&-1,&\dots,-1&\Big) \\
    C_{k'= K-1}      &= \Big(& 0,&  1,& 1,&\dots, 1&\Big) \\
  \end{array}\,. 
\eeq%

\subsubsection{Hyper-systolic matrix multiplication}

The hyper-systolic matrix multiplication, as given in \alg{HYPER-SYSTOLIC},
proceeds within three steps.  In the first part of the algorithm,
matrix $\matrix{B}$ is shifted $K-1$ times by strides of $1$ along the
systolic ring and stored as $B^{i}$, $0\le i\le K-1$. However, as
motivated above, for the case of matrix products, we can spare
communication: it suffices to shift $\matrix{B}$ in $\tilde K$ row
blocks of $K$ rows each, where within each block the first row is
shifted by a stride of $0$ and the last by a stride of $K-1$.

\begin{algo}{Hyper-systolic matrix multiplication.}%
  {\label{HYPER-SYSTOLIC}}
for\= for\= for\= for\= forforfor\= forforforfor\= \kill  
foreach processor $i=1:p$ $\in$ systolic array                  \\
\> \\
\> for $j=1:p$ \> \> \> \> \> ! {\em pre-shift of matrix $\matrix{B}$} \\
\> \>  $b_{j,i} = \cs{[(1-j)\mod{K}]}{p}\big(b_{j,i}\big)$    \\
\> end for                         \\
\\
\> for $j=1:\tilde K-1$ \> \> \> \> \>! {\em multiplication and shift of matrix $\matrix{A}$}  \\
\> \> for $l=1:K$                 \\ 
\> \> \> for $n=1:p$    \\ 
\> \> \> \> $c^{l}_{n,i}=c^{l}_{n,i}+
\cc{[1-(j-1)K-l]}{p}\big(a_{n,i}\big)\,b_{[(j-1)K+l],i}$  \\
\> \> \> end for                  \\ 
\> \> end for \\   
\> \> for $l=1:p$                          \\
\> \> \>  $a_{l,i} = \cs{K}{p}\big(a_{l,i}\big)$    \\
\> \> end for \\
\> end for                        \\ 
\> for $l=1:K$                 \\ 
\> \> for $n=1:p$    \\ 
\> \> \> $c^{l}_{n,i}=c^{l}_{n,i}+
\cc{[1-(\tilde K-1)K-l]}{p}\big(a_{n,i}\big)\,b_{[(\tilde K-1)K+l],i}$  \\
\> \> end for                  \\ 
\> end for \\   
\> \\
\> for $j=1:K-1$ \> \> \> \> \>! {\em back shift and accumulation}\\
\> \> for $l=1:p$                    \\
\> \> \>  $c^{K-j}_{l,i}=c^{K-j}_{l,i}+\cs{1}{p}\big(c^{K-j+1}_{l,i}\big)$\\
\> \> end for                     \\
\> end for                        \\
\\
end foreach                       \\ 
\end{algo}

After the preparatory shifts of $\matrix{B}$, the computation starts.
$\tilde K$ times, the multiplication of $\matrix{A}$ with $K$ rows of
the pre-shifted matrix $\matrix{B}$ is carried out. After each step,
$\matrix{A}$ is moved to the left by a shift of stride $K$.  The
result is accumulated within $K$ matrices $\matrix{C}^{i}$.

Finally, the $K$ intermediate result matrices $\matrix{C}^{i}$ are
shifted back according to base $C_{k'}$ while summed up to the final
matrix $\matrix{C}$.  The algorithm is very regular.  The skew order
is not destroyed during execution, and in any stage, only global
addresses are required.

\subsubsection{Complexity}

The gain factor for the matrix product reads (note that matrix
$\matrix{B}$ is only partially shifted):
\begin{theorem}\label{THE:GAIN3}
  The gain factor $R$ that compares the regular hyper-systolic matrix
  multiplication to the systolic algorithm is
\beq%
R=\frac{p-1}{K+\tilde K-1} \approx \frac{\sqrt{p}}{2}.
\eeq%
\end{theorem}
\begin{pf}
  One needs $1$ shift of the full matrix $\matrix{B}$, $\tilde K-1$ shifts
  by $K$ of matrix $\matrix{A}$ and again $K-1$ shifts by $1$ of matrix
  $\matrix{C}$. Therefore, the total number of shifts required is
\beq%
T=(K+\tilde K-1).
\label{EQ:COMPL}
\eeq%
The standard systolic computation requires $p-1$ shifts of the matrix
$\matrix{A}$.  For the $K=\tilde K=\sqrt{p}$, $R\approx
\frac{\sqrt{p}}{2}$. \hfill\qed
\end{pf}

\subsubsection{Comparison to Cannon's algorithm}

In order to compare Cannon's algorithm and the hyper-systolic matrix
multiplication, we consider a 2D $\sqrt{p}\times \sqrt{p}$ processor
array, where Cannon's algorithm is carried out, and a 1D ring array of
$p$ processors where the hyper-systolic matrix product algorithm is
computed.

The total number of shift operations of Cannon's algorithm is
$2\sqrt{p}-2=2K-2$, while the number of shift operations for the
hyper-systolic algorithm was $2K-1$. Thus, the complexities in terms
of circular shift operations of both algorithms are equal for large
$p$.  We will see below how this fact translates into execution times
on mesh and grid based machines.

\section{Mapping on Parallel Systems\label{BLOCKING}}

So far we discussed the generic situation of the matrix dimension $p$
being equal to the number of processors $p$.  We now turn to the
general case of square $n\times m$-matrices with $n,m >p$.

In order to map the systolic system onto the parallel implementation
machine we choose {\em hierarchy mapping} of the systolic array onto
the processors with the option for two different strategies, block and
cyclic assignment.

\subsection{Block Mapping}
The block assignment is applied in all standard algorithms, as it
allows to exploit local BLAS-3 routines, like {\sf dgemm}, by which a
very high efficiency of local computations can be achieved. While a
small block of the matrix $\bmatrix{A}$ is hold in the cache or in the
registers (thus avoiding cache-to-memory data transfer), in turn, only
the columns of $\bmatrix{B}$ and $\bmatrix{C}$ must be exchanged, and
all computations in which the given part of $\bmatrix{A}$ is involved
can be carried out.  In this way, the ratio between the number of
computations and the cache-to-memory traffic is minimized to nearly
\begin{equation}
\frac{2l^{3}}{(3nn+n^{2})}=\frac{l}{2} 
\end{equation}
floating point operations per word for real data, with $l$ being the
dimension of the sub-block. Asymptotically, the full speed of the CPU
should be exploitable. 

A $n\times m$-matrix $\bmatrix{M}$ is divided into $p\times p$ blocks
of size $\left(\frac{n}{p}\times\frac{m}{p} \right)$ or
$\left(\frac{m}{p}\times\frac{n}{p} \right)$,
\begin{equation} 
  \bmatrix{M}\rightarrow \matrix{M}_{i,j},\quad i=1,\dots,p,\quad
  j=1,\dots,p.
\end{equation} 
The multiplication of $\bmatrix{A}$ and $\bmatrix{B}$ proceeds via
sub-matrix multiplication denoted as $(\otimes)$:
\begin{eqnarray}
\matrix{C}_{i,j} &=& \sum_{k=1}^{p} \matrix{A}_{i,k} \otimes
\matrix{B}_{k,j}, \quad i= 1, \dots,p,\quad j= 1,
\dots,p,
\nonumber\\
\bmatrix{C} &=& \bmatrix{A}\bmatrix{B}.  
\end{eqnarray} 

Altogether a system of $p\times p$ of such blocks is assigned to the
$p$ processor array.  Now we can use each sub-matrix in the same
manner as the scalar matrix elements before.  Therefore, the $p\times
p$ system of sub-matrices has to be row-skewed for $\bmatrix{A}$ and
column-skewed for $\bmatrix{B}$.

\subsection{Cyclic Mapping}

Each block is distributed across the processors as described above for
the generic case.  The blocking of the $n\times m$-matrix
$\bmatrix{M}$ into $\left(\frac{n}{p}\times\frac{m}{p} \right)$
blocks, leads to blocks of size $p\times p$,
\beq%
  \bmatrix{M}\rightarrow \matrix{M}_{i,j},\qquad i=1,\dots,\frac{n}{p},\;
  j=1,\dots,\frac{m}{p}.
\eeq%
The multiplication of $\bmatrix{A}$ and $\bmatrix{B}$ proceeds via
block-multiplication, $(\otimes)$:
\begin{eqnarray}
\matrix{C}_{i,j} &=& \sum_{k=1}^{\frac{m}{p}} \matrix{A}_{i,k} \otimes
\matrix{B}_{k,j},\qquad
i= 1, \dots,\frac{n}{p},\quad j= 1,
\dots,\frac{n}{p},
\nonumber\\
\bmatrix{C} &=& \bmatrix{A}\bmatrix{B}.  
\end{eqnarray} 
A skew representation is required for all blocks $\matrix{M}_{i,k}$
separately.  Cyclic mapping leads to a system of $\frac{n}{p}\times
\frac{m}{p}$ systolic processes that run in parallel.

Cyclic assignment allows us to reduce the memory overhead of
hyper-systolic computations.  In general, $K$ full intermediate
matrices $\bmatrix{C}$ are necessary. Using cyclic mapping, one can
organize the computation in such a way that only one row of the blocks
of the intermediate matrix $\bmatrix{C}$ must be stored in a given
phase of the algorithm.  All the required shifts of the given part of
$\bmatrix{A}$ can be carried out while this part of $\bmatrix{A}$ will
not be involved in a further computation.  Eventually, the
corresponding row of $\bmatrix{C}$ is shifted back and accumulated.

A second interesting feature of cyclic mapping is the possibility to
distribute the very blocks. This approach is interesting for machines
with a hybrid architecture like the proposed Italian PQE2000 system
\cite{PQE2000}.  In this approach, the rows of $\bmatrix{A}$ and the
columns of $\bmatrix{B}$ are assigned to different processors each.

\subsection{Block-Cyclic Mapping}

One can combine block and cyclic mapping in a hybrid scheme that
combines the advantages of both approaches.  A good strategy is to
choose the block size of the block mapping such that it is optimal for
``local'' BLAS-3. For the cyclic part one ends up with blocks of size
$p\times p$, with the entries being the BLAS-3 blocks.

\section{Implementation Issues\label{TESTS}}

The class of algorithm presented can be useful on any type of
massively parallel system with distributed memory. Mesh and grid-based
connectivities  might benefit as well as large
work-station clusters.

In the previous section, we have presented complexities of
interprocessor communication in terms of circular shifts, irrespective
of the actual communication time of a shift. This time is a constant
on workstation clusters usually connected via Ethernet, therefore
complexities in terms of circular shifts will translate into real time
in a straightforward way. On mesh and grid based machines 1D rings in
general can be realized as a subset of their system connectivity.

\begin{figure}[b]
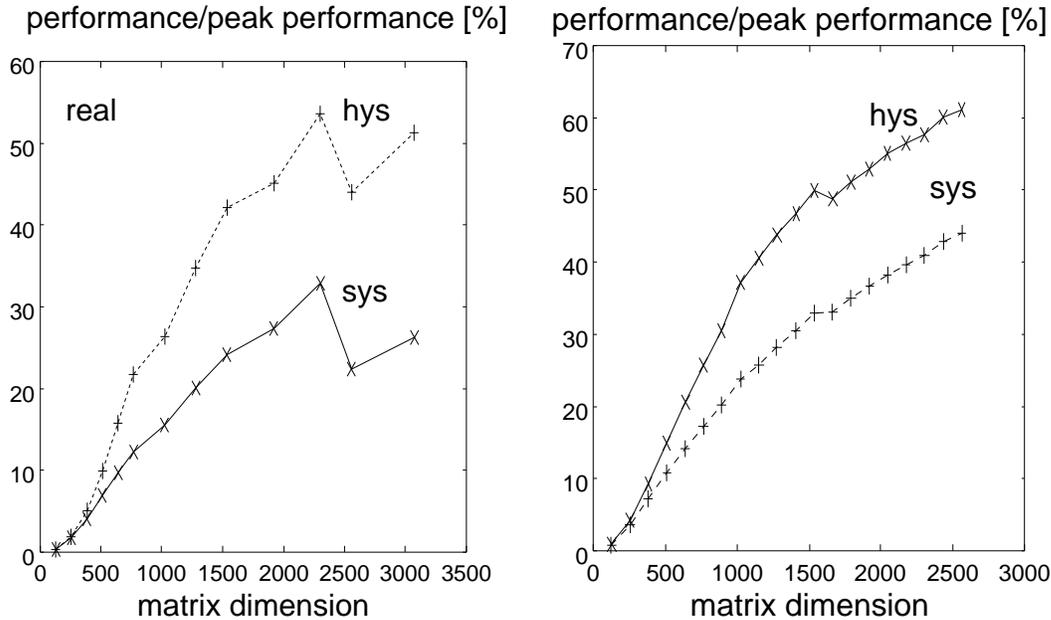

\begin{minipage}{\textwidth}
\centerline{\epsfxsize=.48\textwidth\epsfbox{perf_r.eps}\hfill
\epsfxsize=.48\textwidth\epsfbox{perf_c.eps}}
\end{minipage}
\caption[Performance Measurements of the Matrix Product]{
  Performance of systolic and hyper-systolic PBLAS-3 (block-cyclic
  mapping) on a 128-node APE100, for real and complex data.
\label{FIG:PERFQH1_B}}
\end{figure}

As an example taken from real life, we have implemented a level-3
PBLAS code on the APE100 parallel computer.  So far, on APE100 lack of
indexed addressing has hindered an effective scalable implementation
of PBLAS \cite{VICERE}.  In our approach, we made use of a combination
of block and cyclic mapping. Thus, we are able to use local BLAS,
exploiting the CPU with high efficiency, and to reduce the memory
overhead of the hyper-systolic algorithm.

On APE100, for real data, the optimal elementary blocks are of size
$6\times 6$. For complex data, the size is $4\times4$.  The full
matrix is blocked to $p\times p$ matrices which are distributed on the
ring and the elements of which are the elementary blocks.  Only the
$p\times p$ matrices are skew, the elementary matrices remain in
normal order.

We have benchmarked a 128-node APE/100Quadrics QH1 using real and
complex data.  \fig{FIG:PERFQH1_B} shows the performance results.

The theoretical peak performances (single node!) for Quadrics are $63$
\% for real data and $88$ \% for complex data, as can be inferred from
the maximal ratio of computation \vs\ memory-to-register data transfer
times. Hyper-systolic matrix multiplication leads to a peak
performance of 65 \% of peak speed, which translates into 75 \% of the
theoretical performance.

\section{Summary}

The 1D hyper-systolic matrix multiplication algorithm is a promising
alternative to 2D matrix product algorithms. With equal complexity as
standard methods, the hyper-systolic algorithm avoids non-regular
communication and indexed local addressing. Hence, the hyper-systolic
matrix product scheme is universal: it is applicable on any type of
parallel system, even on machines that cannot compute local addresses.
The method preserves the alignment of the matrices in the course of
the computation. Besides the fact that transposed matrix products can
be carried out on the same footing, alignments for the optimal
hyper-systolic algorithm are efficient for matrix-vector
operations as well.

So far, in a feasibility study, the hyper-systolic matrix product has
been implemented on APE100/Quadrics systems \cite{PDPTA97,BLAS97}. We
will present details of the implementation on this system for matrix
and transposed matrix products in Ref.~\cite{CETRARO2}.


\begin{thebibliography}{99}
\frenchspacing
%
\bibitem{DONGARRA} J. Choi, J. J. Dongarra, and D. W. Walker:
``The Design of Scalable Software Libraries for Distributed Memory
Concurrent Computers'',  in: J. J. Dongarra and B. Tourancheau (eds.):
{\it Environments and Tools for Parallel Scientific Computing}
(Elsevier, 1992).
%
\bibitem{LIPPERT95} Th. Lippert, A. Seyfried, A. Bode, K. Schilling:
  `Hyper-Systolic Parallel Computing',  IEEE Trans. on Parallel
    and Distributed Systems 9 (1998) 1.
%
  \bibitem{GALLI95} A. Galli: `Generalized Hyper-Systolic Parallel
    Computing', preprint server hep/lat,
    URL:~http://xxx.lanl.gov/ps/hep-lat/9509011
%
\bibitem{FOX87} G. C. Fox and S. W. Otto: `Matrix Algorithms on a
  Hypper-Cube I: Matrix Multiplication', Parallel Computing 4 (1987)
  17-31.
%
\bibitem{BOOK1} N. Petkov: {\it Systolische Algorithmen und Arrays}
  (Berlin: Akademie-Verlag, 1989).
%
\bibitem{BOOK2} N. Petkov: {\it Systolic Parallel Processing}
  (Amsterdam: North-Holland, 1993).
%
\bibitem{KUNG1} H.T. Kung and C. E. Leiserson: ``Systolic arrays (for
  VLSI)'', {\it Sparse Matrix Proc.} 1978 (Society for Industrial and
  Applied Mathematics, 1979) pp.256-282; the same as ``Algorithms for
  VLSI processor arrays'', in C. Mead and L. Conway: {\it Introduction
    to VLSI Systems} (Reading, MA: Addison-Wesley, 1980) sect.8.3.
%
\bibitem{KUNG2} H.T. Kung: ``Why systolic architectures'', 
    Computer 15 (1981) pp.37-47.
%
  \bibitem{CAPPELLO83} P.R. Cappello and K. Steiglitz: ``Unifying VLSI
    design with geometric transformations'', Proc. Int. Conf.
    Parallel Processing (1983) 448-457.
%
  \bibitem{CAPPELLO87} P.R. Cappello: ``Space time transformation of
    cellular algorithms'', in E.E. Swartzlander (ed.), {\it Systolic
      Signal Processing Systems} (N.Y., Basel: Dekker, 1987), 161-208.
%
\bibitem{QUINTON84} P. Quinton: ``Automatic synthesis of systolic
  arrays from uniform recurrent equations'', {\it Proc. 11th Annual
    Int. Symp.  Comput.  Archit.}, Ann Arbor, Mich., 1984 (IEEE, N.Y.,
  1984) pp.  208-214.
%
\bibitem{MOLDOVAN82} D.I. Moldovan: "On the analysis and synthesis of
  VLSI algorithms", IEEE Trans. on Computers C-31 (1982) 1121-1126.
%
\bibitem{MOLDOVAN83} D.I. Moldovan: ``On the design of algorithms for
  VLSI systolic arrays'', Proc. IEEE 71 (1983) 113-120.
%
\bibitem{CLAUSS94} P. Clauss, G. R. Perrin: ``Optimal Mapping of
  Systolic Algorithms by Regular Instruction Shifts'', IEEE
  International Conference on Application-Specific Array Processors,
  ASAP (1994) 224-235.
%
\bibitem{DARTE95} A. Darte, Y. Robert: ``Affine-by-Statement
  scheduling of uniform and affine loop nests over parametric
  domains'', Journal of Parallel and Distributed Computing, 29 (1995)
  43-59.
%
\bibitem{CLAUSS96} P. Clauss , V. Loechner, ``Parametric analysis of
  polyhedral iteration spaces'', IEEE International conference on
  Application Specific Array Processors, ASAP, 1996.
%
\bibitem{DONTJE91} T. Dontje, Th. Lippert, N. Petkov and K. Schilling:
  ``Statistical analysis of simulation-generated time series: Systolic
  vs.  semi-systolic correlation on the Connection Machine'', Parallel
  Computing 18 (1992) 575-588.
%
\bibitem{PETKOV_FUZZY} N. Petkov: `Fuzzy number subtraction
  convolution on the CM-2', Int. J. of Mod. Phys. C4 (1993) 181-196.
%
\bibitem{PETKOV_FUZZY2} N. Petkov: `Fuzzy number subtraction
  convolution on the CM-2', in: Th. Lippert, K.  Schilling and P.
  Ueberholz (eds.) {\it Science on the Connection Machine} (Singapore:
  World Scientific, 1993), pp. 181-196.
%
\bibitem{LIPPERT96} Th. Lippert, U. Glaessner, H. Hoeber, G.
  Rit\-zen\-h\"ofer, K. Schilling, and A. Seyfried: `Hyper-Systolic
  Processing on APE100/Quadrics, I. $n^2$-loop computations',
    Int.  Jour. Mod. Phys. C 7 (1996) 485.
%
\bibitem{HEGGIE} G. Meylan and D.  C. Heggie: `Internal Dynamics of
  Globular Clusters', {\it preprint HEP-ASTRO,
    http://xxx.lanl.gov/multi}, in press in {\it The Astronomy and
    Astrophysics Review}.
%
\bibitem{LIPPERTPHD} Th. Lippert: `Hyper-Systolic Parallel
  Computing---Theory and Applications', PhD-thesis, University of
  Groningen, 1998.
%
\bibitem{EKLUND} J. O. Eklundth: `A fast Computer Method for Matrix
  Transposing', IEEE Trans. on Computers C 21 (1972) 801-803.
%
\bibitem{DEPREZ} F. DePrez and M. Pourzandi: `A Comparison of three
  Parallel Matrix Product Algorithms', {Proc. of the Int. Conf.
    Advances in Numerical Methods \& Applications, Sofia} (1994)
  234-244.
%
\bibitem{PARCO95} H. Gupta and P. Sadayappan: `Communication Efficient
  Matrix Multiplication on Hypercubes', Parallel Computing 22
  (1996) 75-99.
%
\bibitem{CAN69} L. E. Cannon: `A Cellular Computer to Implement the
  Kalman Filter Algorithm', PhD Thesis, Montana State University,
  1969.
%
\bibitem{KUMAR} V. Kumar, A. Grama, A. Gupta, and G. Karypis: {\it
    Introduction to Parallel Computing} (Redwood City:
  Benjamin/Cummings, 1994).
%
\bibitem{HPF} `High Performance Fortran Language Specification,' Rice
  University, version 1.1 November 1994.  `High Performance Fortran',
  {\it Scientific Programming}, 2 (1993).
%
\bibitem{HOFMEISTER} M. Djawadi and G. Hofmeister: `The Postage Stamp
  Problem', {\it Mainzer Seminarberichte}, Additive Zahlentheorie 3
  (1993) 187.
%
\bibitem{UNSOLVED} R. K. Guy, {\it Unsolved Problems in Number
    Theory}, (Springer-Verlag, Berlin, New-York, 1994).
%
\bibitem{PQE2000}
  URL:(http://www.sede.enea.it/~hpcn/moshpce/hpcn01e.html)
%
\bibitem{VICERE} M. Beccaria, G. Cella, A. Ciampa, G. Curci, and A.
  Vicer\'e: `Matrix Inversion on APE100 Machines',  Preprint
    IFUP-TH 17/95.
%
\bibitem{PDPTA97} Th. Lippert and K. Schilling: `Hyper-Systolic Matrix
  Multiplication', in: H. R. Arabnia (ed.), {\it Proceedings of the
    International Conference on Parallel and Distributed Processing
    Techniques and Applications, PDPTA '96}, Sunnyvale, California,
  USA, - August 9 - 11, 1996, (CSREA, 1996), pp. 919-930.
%
\bibitem{BLAS97} Th. Lippert, N. Petkov, and K. Schilling: `BLAS-3 for
  the Quadrics Parallel Computer', in: B,. Hertzberger and P. Sloot
  (eds.), {\it Proceedings of the International Conference on High
    Performance Computing and Networking, HPCN '97}, Vienna, Austria,
  April 1997, (Springer, Berlin, 1997) pp. 332-341.  919-930.
%
\bibitem{CETRARO2} M. Coletta, Th. Lippert, P.Palazzari, N. Petkov,
  and K. Schilling: to appear.
%
\nonfrenchspacing
\end{thebibliography}
\end{document}